# Organometallic Complexes of Graphene and Carbon Nanotubes: Introducing New Perspectives in Atomtronics, Spintronics, High Mobility Graphene Electronics and Energy Conversion Catalysis

**Santanu Sarkar,*$ and Robert C. Haddon**

**Center for Nanoscale Science and Engineering, Departments of Chemistry and Chemical & Environmental Engineering, University of California, Riverside, CA 92521-0403, United States**
*E-mail: ssark002@ucr.edu

$Present address: Intel Corporation, Logic Technology Development, Ronler Acres Campus, Hillsboro, OR 97124, United States
*E-mail: santanu.sarkar@intel.com

**ABSTRACT:** Here we present an overview of recent fundamental studies on the nature of the interaction between individual metal atoms and metal clusters and the conjugated surfaces of graphene and carbon nanotube with a particular focus on the electronic structure and chemical bonding at the metal−graphene interface. We discuss the relevance of organometallic complexes of graphitic materials to the development of a fundamental understanding of these interactions and their application in atomtronics as atomic interconnects, high mobility organometallic transistor devices, high-frequency electronic devices, organometallic catalysis (hydrogen fuel generation by photocatalytic water splitting, fuel cells, hydrogenation), spintronics, memory devices and in the next generation energy devices. We touch on CVD graphene grown on metals, the reactivity of its surface, and its use as a template for asymmetric graphene functionalization chemistry (ultrathin Janus discs). We highlight some of the latest advances in understanding the nature of interactions between metals and graphene surfaces from the standpoint of metal overlayers deposited on graphene and SWNT thin films. Finally, we comment on the major challenges facing the field and the opportunities for technological applications.

# 1. INTRODUCTION

Graphene possesses a 2D gas of confined electrons with a singular electronic structure at the Dirac point, and its ultrahigh electronic mobility, and potential in nanoelectronics, energy devices, bionanotechnology, single-molecule sensing, and advanced energy devices has stimulated strong interest in this material.[1] While the fundamental physics of graphene has already been demonstrated,[2-5] the exciting chemical implications of the massless two-dimensional gas of Dirac fermions in graphene is just now coming into focus.[6, 7] The recent exploration of the chemistry of graphene at the Dirac point has provided a rational understanding of the chemical reactivity of graphene in Diels-Alder pericyclic reactions, based on the graphene frontier molecular orbitals at the Dirac point as they relate to the original frontier molecular (FMO) theory and orbital symmetry conservation concepts.[6, 8-11] This understanding has provided a unified theory of graphene reactivity, including radical addition chemistry, Diels-Alder pericyclic chemistry, and organometallic mono- and bis-hexahapto metal complexation chemistry.[12, 13] Nonetheless, the future applications of graphene in carbon-based electronics require (1) high quality electrical contacts to graphene, (2) introduction of a band gap (semiconducting behavior) in the zero-band-gap semi-metallic graphene, and (3) the production of high quality large-area graphene wafers, which will allow standard wafer-scale lithographic patterning and etching for scalable device fabrication.[14] While the 2D nature of graphene is compatible with standard organic chemistry processing and lithographic patterning of graphene wafers, defining high quality metal contacts to graphene calls for an in-depth understanding of the conditions necessary for the growth of uniform metal films (by e-beam evaporation or sputtering deposition) and the nature of metal-graphene interfaces at a fundamental level.[15] Additionally, the fundamental understanding of the interaction between mobile metal atoms or metal nano-clusters and graphitic surfaces is crucial from the standpoint of CVD growth of graphitic materials on metal surfaces (surface catalysis),[16] spintronics (spin filters),[17] electronic devices (ultrafast graphene transistors, memory devices),[17] atomic interconnects,[18-21] and superconducting phenomena.

At this point it is clear that 2D graphene is in some ways a simpler version of the surfaces offered by the 1D single-walled carbon nanotubes which have attracted the attention of chemists, material scientists and physicists since the advent of the 0D fullerenes. In this Perspective we briefly cover the interaction of metals with the graphitic surfaces of all of these materials with an emphasis on the covalent bonding approach introduced in our recent research on carbon nanotubes and graphene.[6, 12, 13]

Recently, we have explored the organometallic hexahapto (metal) complexation chemistry of graphene (epitaxial graphene, CVD graphene, and exfoliated graphene flakes), graphite (HOPG and graphite nanoplatelets), and single-wall carbon nanotubes,[18-23] in which the benzenoid surfaces of the graphitic materials act as the active ligand. We have found that metal atoms provide a conductive pathway between two adjacent SWNTs, in which the metal atom forms a bis-hexahapto-bridge thereby functioning as highly conductive ($\eta^6$-SWNT)M($\eta^6$-SWNT) interconnects and significantly decreasing the inter-nanotube junction resistance.[18-20] The ability to electrically interconnect benzenoid surfaces is an important step toward the fabrication of molecular and single layer solid state carbon-based circuitry.[24] Transition metals from Groups 4, 5, 6, 7 and 8 were found to function as an electrical conduit between SWNTs via the constructive overlap between the metal $d_\pi$ and SWNTs $\pi$-orbitals. The canonical analogue of this type of carbon-based interconnect is bis(benzene)chromium [($\eta^6$-benzene)$_2$Cr], which is well known in chemistry since its first synthesis by Hafner and Fischer.[25] Our conductivity experiments utilized high vacuum electron-beam evaporation of metals onto SWNT thin films,[18-20] in analogy with the metal vapor synthesis (MVS) technique used in many of the original preparations of the ($\eta^6$-benzene)$_2$M complexes.[26-28]

Recently it has been shown that rare-earth (RE) metal islands (composed of dysprosium, gadolinium, and europium) on graphene act as tiny magnets, with very high density.[17, 29, 30] and these graphene–metal structures have been suggested to have application in memory device architectures.[17]

Conventional bulk metal contacts are known to dope graphene,[31] and this can be used to controllably adjust the Fermi level without disturbing the electronic structure of graphene at the Dirac point.[32] The Schottky barrier at the metal electrode/graphene channel interface is determined by the energy difference between the metal work function (W) and the electron affinity of the active channel. Electron injection into the conduction band of the graphene channel can be achieved by using metal electrodes with low work function whereas metal electrodes of high work functions are adequate for hole injection into the valence band.[32] The systematic variation of the work function of the metal electrode may provide an approach to precisely tailoring the electrical properties of the device, thereby modifying the energy level alignment near the electrode/graphene or SWNT interface.[32, 33] Knowledge of the exact nature of the graphene–metal and SWNT-metal interactions is crucial for an understanding of the doping (ionic charge transfer) to graphene and SWNTs by metals and the possibility of covalent modification.

Magnetic film growth has also been of interest for spintronics, either for growing continuous uniform films as spin filters or as isolated magnetic atoms with a modified Kondo effect because of the 2D nature of graphene.[17] It has been predicted that novel properties should emerge after transition metal adsorption and that graphene should become magnetic.[34] Additionally, surface magnetochemistry in the novel ($\eta^6$-graphene)Cr(ligand) organometallic systems, in which metal atoms are trapped in the 2D potential lattice of a graphene–ligand interface, is suggested to "open up the possibility in building a chemical analogue of an optical lattice, a key setup in quantum information and strongly correlated systems".[35] Based on theoretical calculations a "general principle of spin–charge separation in π–d systems" was elucidated which suggested that "ligands can work as a local gate to control the properties of trapped metal atoms and can impose bosonic or fermionic character on such atomic nets, depending on the ligand's nature".[35]

In this Perspective, we discuss recent experimental efforts to understand the bonding and electronic structure at the graphene- and SWNT–metal interfaces from the standpoint of graphene growth on metal thin film surfaces as well as on the metal overlayers deposited (by e-beam evaporation or magnetron sputtering) on graphene and SWNT structures. We also provide a perspective on the potential applications of these organometallic complexes of SWNTs and graphene in catalysis (reusable solid support for the water splitting reaction for hydrogen fuel generation, polymer electrolyte membrane fuel cells,[36] and for hydrogenation[37]), in producing Janus graphene-based materials (the possibility of asymmetric two-dimensional graphene chemistry or anisotropic functionalization),[15] high mobility organometallic transistor devices,[23] and 2D and 3D electrical interconnects with high current carrying capacity.[18]

**2. Nature of Interactions between Metals and Graphitic Surfaces.** There are two limiting cases for the interaction of a metal with a graphene surface – that which involves metal atoms and that which involves the bulk metal. In the former case the metal atoms are usually added to the intact graphene surface by either physical or chemical means. In the latter case, where a bulk metal is involved, the graphene is often transferred to the metallic surface or grown directly. The present Perspective will mainly focus on a consideration of metal atoms, clusters and nanoparticles as this reflects current interests within our own research group, but we acknowledge the great importance of the interaction of bulk metals and graphene surfaces, particularly is it relates to the CVD growth of graphene and in defining bulk metal contacts to graphitic surfaces. We will distinguish between four limiting cases for the interaction of metal atoms with graphene surfaces:

(a) Weak physisorption of metal atoms generally occurs when the metal atom has its d-orbitals filled (in the case of transition metals such as gold) or possesses an s,p-like metallic structure with free-electron-like parabolic band structure (such as Pb), together with a high work function.

(b) Ionic chemisorption is characteristic of the interaction of metals of low ionization energy such as alkali metals (Li, Na, K) and alkaline earth metals (Be, Ca. Mg). Metals with low work function lead to the injection of electrons into the conduction band of graphitic materials (n-type doping) whereas metals with higher work function lead to the injection of hole into the valence band (p-type doping). Such charge transfer interaction with the graphitic structure largely preserves the conjugation and band structure of the graphitic system.

(c) Covalent chemisorption of metals to graphitic systems leads to strong (destructive) rehybridization of the graphitic band structure. One such example is the formation of metal carbides by the strong interaction between graphitic surface and metals leading to metal−carbon bond formation (Ti).

(d) Covalent chemisorption of metals to graphitic systems, which is accompanied by the formation of an organometallic hexahapto($\eta^6$)-metal bond, preserves the graphitic band structure (constructive rehybridization), and this provides a distinct type of interaction between metals and graphitic surfaces. We have recently discovered that the constructive rehybridization that accompanies the formation of bis-hexahapto-metal bonds, such as those in ($\eta^6$-SWNT)Cr($\eta^6$-SWNT), interconnects adjacent graphitic surfaces and significantly reduces the internanotube electrical junction resistance in single-walled carbon nanotube (SWNT) networks.[18-20] In the traditional carbon forming covalent chemistry of graphene, the $sp^2$ hybridized carbon atoms at the sites of covalent attachment of functional groups are converted into $sp^3$ centers, which can introduce a band gap into graphene, influence the electronic scattering, and create dielectric regions in a graphene wafer with drastically reduced device mobility.[6, 7, 38] However, the organometallic hexahapto ($\eta^6$) functionalization of the two-dimensional (2D) graphene π-surface with transition metals, does not bring about significant structural rehybridization of the graphitic surface, and provides a new way to modify graphitic structures that does not saturate the functionalized carbon atoms and, by preserving their structural integrity, maintains the delocalization in these extended periodic π-electron systems and offers the possibility of three-dimensional (3D) interconnections between adjacent graphene sheets.[13]

**3. Mobile Metal Atoms on Graphene Surfaces.** A number of surface science tools have been used to image metals on graphene surfaces and some of these techniques provide information on the mobility of the atoms. Angle resolved photoemission spectroscopy (ARPES),[39] near-edge X-ray absorption fine structure (NEXAFS) spectroscopy,[15] scanning tunneling microscopy (STM),[17, 40] scanning transmission electron microscopy (STEM),[41] Raman spectroscopy[42] as well as surface-enhanced Raman spectroscopy (SERS)[43] and first-principles density functional theory (DFT)[15] are generally employed to understand the interaction between metals and graphene-like surfaces. ARPES studies have mapped out the expected linear energy (E) versus wavevector (k) dispersion of graphene at the Dirac point.[39] Near-edge X-ray absorption fine structure (NEXAFS) spectra at the carbon K-edge are sensitive to changes in the bonding environment of graphene and was employed to examine the details of the nature of chemical bonding at the metal-graphene interfaces.[15] In the NEXAFS spectra a sharp splitting of π* resonance implies substantial wavefunction overlap between the graphene π cloud and transition metal d orbitals as a result of hybridization-induced symmetry breaking of the two-atom graphene unit cell and charge redistribution across the interfacial region.[15] With the discovery of the excellent electronic and mechanical properties of suspended (free-standing) graphene devices, which is devoid of the influence from the underlying substrate, many surface scientists have focused their attention on the crystallographic structure and electronic properties of the graphene-metal systems by STM. Low-energy electron diffraction (LEED) and Auger electron spectroscopy (AES) are also employed to study the graphene-metal systems. The strength of interaction between metallic substrates and graphene layers has been studied by high resolution electron-energy loss spectroscopy (EELS).[44]

The interaction of metal atoms on epitaxial graphene (EG on SiC) surfaces has been studied by depositing sub-monolayers of metal atoms on graphene, and the geometry and thermal stability of the self-assembled metal atoms on graphene was observed via scanning tunneling microscopy (STM) and these studies have shown that rare-earth (RE) metals, such as dysprosium (Dy) and gadolinium (Gd), interact strongly with graphene, while lead (Pb) shows weak interaction with graphene.[17, 29] Efforts have been directed towards quantifying the interaction by calculating the structure, bonding, and charge transfer with different metals.[45] The nature of the absorbed metals dictates the relevant energetic parameters like adsorption energy ($E_a$), diffusion energy barrier ($\Delta E$), and charge transfer,[29] and the bonding to the metal was inferred to vary from weak/ionic to strong/covalent. On the basis of the STM experiments, the nucleated island density, the thermal island stability, and crystallization temperature were determined, which were in turn correlated to $\Delta E$, $E_a$, atom detachment, and surface energies in conjunction with other parameters such as charge transfer and electronic dipole moment to clarify the metal–carbon bonding.[45] Subsequent studies supported the classification in terms of weakly interacting free-electron metals and strongly interacting rare earths that lead to significant rehybridization of the graphene surface.[29]

Lead (Pb) possesses s and p valence electrons and shows a free-electron-like parabolic band structure. STM studies have shown that Pb atoms on graphene have long diffusion lengths so that Pb atoms moved quickly on the graphene surface even at temperature as low as 30K forming flat-top Pb(111) islands with intact regions between the layers (**Figure 1a,b**).[29] The charge transfer to graphene and lead-graphene bonding were found to be extremely weak;[46] the reported values of the diffusion barrier (activation energy) were less than 35 meV and 70 meV at 30 K and 78 K, respectively.[17]

The dysprosium (Dy) atom has s, p, and d orbitals in the valence shell with a low lying unfilled f shell. As a rare earth (RE) metal, Dy is far more reactive and shows non-parabolic bands,[29] and STM studies have shown that the room temperature diffusion length of Dy is less than 1/10 of the diffusion length of Pb at 30 K. The dysprosium atoms move slowly, even after heating, indicating strong interaction with the graphene and crystallization occurs only at a much higher temperature (> 660 K).(**Figure 1c,d**).

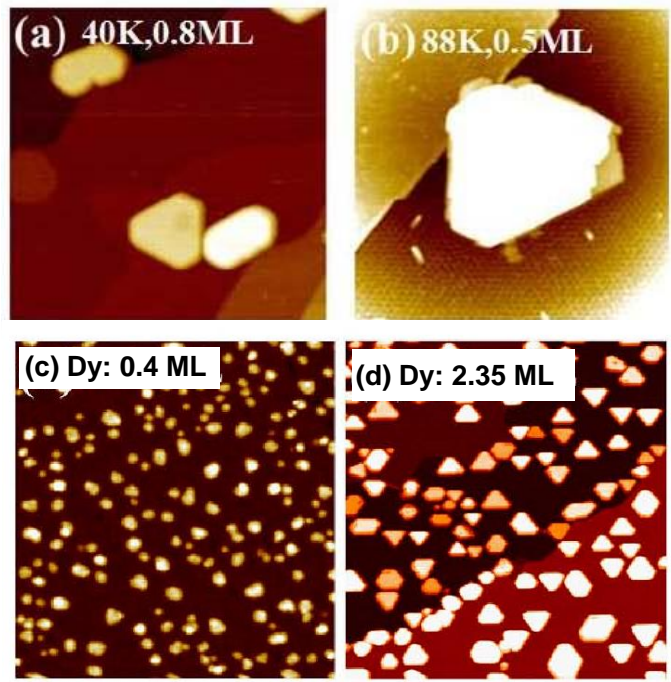

**Figure 1.** (**a**) 250 × 250 nm² area with only 3 large Pb islands the temperature is T = 40 K at a flux rate 0.1 ML/min and deposited amount θ = 0.8 ML. The measured island density is $3 \times 10^{-5}$ islands/nm² implying a Pb diffusion barrier less than 50 meV. (**b**) The area is 54 × 44 nm² Pb after growth at T = 88 K and θ = 0.5 ML. The periodic pattern around the 7-layer island is the initial 6√3 reconstruction which is practically free of monomers indicating unusually high Pb mobility. (c) The nucleated island density after 0.4 ML of Dy has been deposited at room temperature over an area of 1000 × 1000 Å². The island density is $2 \times 10^{-4}$ islands/Å² indicating slow diffusion. (d) Dy (2.35 ML) deposited on graphene with a flux rate 0.23 ML/min at 660 K. The area size is 2500 × 2500 Å². These islands have height difference multiples of 0.28 nm, the Dy step height. Reprinted from ref.[29] with permission by John Wiley and Sons. Copyright 2011 WILEY-VCH Verlag GmbH & Co. KGaA, Weinheim.

Europium (Eu) deposition on epitaxial graphene (EG) surfaces quickly results in flat films with large terraces; the deposition of 1 ML of Eu on graphene at RT (deposition rate 0.2 ML/min) showed large and crystalline islands, principally of height 0.75 nm with well-defined facets (**Figure 2d**). The film grows in a layer-by-layer fashion if the deposition of Eu continues beyond 3 ML (**Figure 2e**).[17]

Gadolinium (Gd) on graphene showed stronger interaction; the room temperature (RT) deposition of Gd at a flux rate of F = 0.12 ML/min showed fractal-like Gd 3D islands at all coverages (**Figure 2f-h**). While the island size increases with coverage (θ), the island density *n* is constant at approximately 0.0014 islands/nm² over a wide range of coverage.[17]

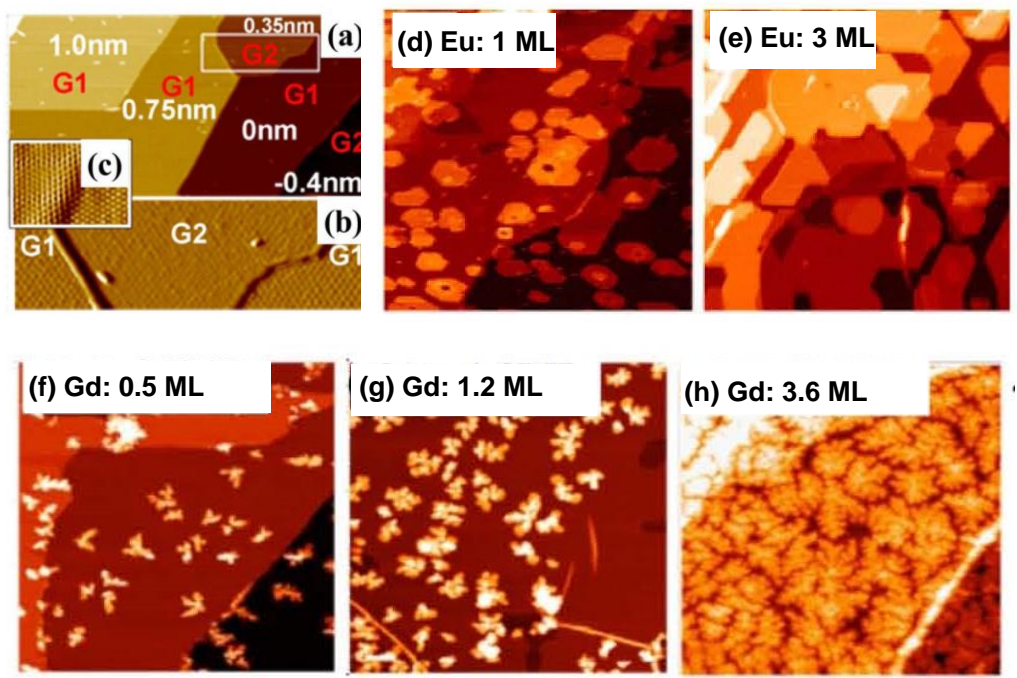

**Figure 2.** (**a**) 250 × 250 nm² area scanning tunneling microscopy (STM) image showing different heights on a graphitized 6H-SiC(0001) sample; (**b**) The outlined white box of Figure 5a. All step heights can be written in terms of s = 0.25 nm or g = 0.35 nm; (**c**) Single layer graphene growing uninterrupted over a graphene step. (d) 250 × 250 nm² STM images for θ= 1 ML of Eu deposited on graphene at RT with a deposition rate 0.2 ML/min. The islands are large and crystalline of main height 0.75 nm with well-defined facets; (e) The film closes in a layer-by-layer fashion if the deposition of Eu continues to 3 ML as shown in the STM image of 250 × 250 nm². Panel (a-e) are reprinted from ref. [17]. Copyright 2013 MDPI AG, Basel, Switzerland. (f-h) STM image of Gd on graphene (f) 210 × 210 nm², θ = 0.5 ML, (g) 250 × 250 nm², θ = 1.2 ML; (h) 250 × 250 nm², θ = 3.6 ML. Panel (f-h) are reprinted from ref.[30]. Copyright 2011 IOP Publishing Ltd.

From the standpoint of organometallic chemistry, if a chromium (Cr) atom is bonded in hexahapto ($\eta^6$) fashion to one of the graphene benzenoid rings, the complex is six electrons short of the stable 18-electron electron configuration.[22] Chromium atoms are mobile on graphitic surfaces,[22, 41] and we found that Cr atoms on SWNTs promptly move to a carbon nanotube (CNT) junction to coordinate to the benzenoid rings of another SWNT so as to obtain the stable 18-electron configuration of ($\eta^6$-arene)$_2$Cr (where arene = SWNT).[18-20] A number of other transition metals such as Ti interact strongly with the graphene surface and this results in decreased mobility.[45, 47] In contrast to these transition metals which strongly interact with the graphene surface, gold interacts weakly and the strength of the Au-Au interaction leads to ready cluster formation.[41, 48]

## 4. ATOMTRONICS

The electrical connection of graphitic surfaces to bulk metal wiring constitutes a major problem in most approaches to molecular electronics, individual carbon nanotube devices or graphene circuitry.[32, 49, 50] In an attempt to address this problem in materials with graphitic surfaces – carbon nanotubes, graphenes and other forms of benzenoid-based carbon materials - we have used single atom bridges to develop a technology we term atomtronics. Below we discuss the application of atomtronics to electrically connect graphene surfaces via bis-hexahapto-metal complexation reactions.

**4.1. High Mobility Organometallic Graphene Transistors via Mono-Hexahapto ($\eta^6$)-Metal Complexation.** We recently reported the organometallic hexahapto ($\eta^6$)-chromium metal complexation of single-layer graphene (SLG) to produce field effect transistor (FET) devices which retain a high degree of the mobility and show an enhanced on-off ratio (**Figure 3**).[22, 23] This $\eta^6$-mode of bonding is quite distinct from the modification in the electronic structure induced by conventional covalent σ-bond formation which involves the creation of $sp^3$ carbon centers in the graphene lattice.[6, 7, 11, 12, 38] Thus the application of organometallic functionalization chemistry has enables the fabrication of FET devices which retain the high mobility of graphene, presumably due to the fact that such organometallic hexahapto functionalization preserves the conjugation of these extended periodic π-electron systems and the functionalized carbon atoms remain a part of the electronic band structure. In other words, the degree of rehybridization at the site of complexation is insufficient to saturate the conjugated electronic structure, unlike those reactions that require destructive hybridization,[38] which when incorporated in electronic field effect devices show low conductivity and significantly reduced carrier mobility.[13]

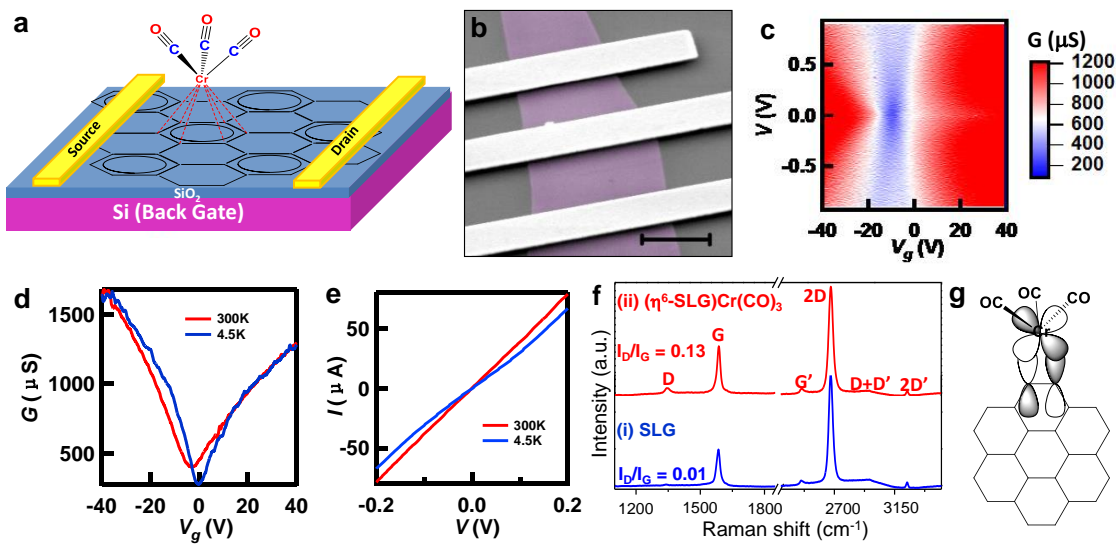

**Figure 3.** Fabrication of high mobility organometallic graphene transistor devices. (a) Schematics of a ($\eta^6$-SLG)Cr(CO)$_3$ organometallic single-layer graphene (SLG) FET device on an oxidized silicon wafer with metal contacts (Au/Cr = 150 nm/10nm). (b) False-color scanning electron microscopic (SEM) of a typical graphene device (scale bar 2 µm), with color of graphene (SLG) matching to that seen in optical microscope. (c) Conductance $G$ as a function of bias $V$ and gate $V_g$ at 4.5 K of a ($\eta^6$-SLG)Cr(CO)$_3$ organometallic device. (d,e) $G(V)$ characteristics and $I(V)$ curves of a weakly functionalized device at 300 K and 4.5 K. The functionalized graphene device has mobility of ~2,000 cm$^2$V$^{-1}$s$^{-1}$ at room temperature and ~3,500 cm$^2$V$^{-1}$s$^{-1}$ at 4.5 K. (f) Comparison of Raman spectra ($\lambda_{ex}$ = 0.7 µm) of a (i) pristine SLG and (ii) metal functionalized devices, which shows a small increase in $I_D/I_G$ (from 0.01 to 0.13). (g) Schematic illustration of the fact that the flat two-dimensional structure of graphene makes it an ideal hexahapto ligand. Adapted from ref.[23] with permission by John Wiley and Sons. Copyright © 2013 WILEY-VCH Verlag GmbH & Co. KGaA, Weinheim.

**4.1.1. Solution Phase Synthesis of Organometallic Complexes of Graphene.** The ($\eta^6$-graphene)Cr(CO)$_3$ complexes are generally synthesized by thermolysis of Cr(CO)$_6$ under an inert atmosphere in the presence of high-boiling-point solvents.[22, 51] The solvent can be dibutylether/THF, 1,2-dimethoxyethane (DME), diglyme/THF, heptane/diglyme, α-picoline, decalin, decalin/ethyl formate or decalin/butyl acetate.[51] The polar ether or ester additives (or solvents) promote carbonyl dissociation, stabilize the intermediates, and the vigorous reflux of lower boiling additives (such as THF) washes the sublimed Cr(CO)$_6$ from the reflux condenser back into the reaction mixture. The same metal complexation reaction can also be performed at lower reaction temperature using this procedure, but with the presence of naphthalene (~0.25 equivalents) as an additional ligand in the reaction mixture.[23] This is due to the in-situ formation of the (naphthalene)Cr(CO)$_3$ complex, which is highly labile due to the localization of the π-electrons, thus facilitating arene exchange reactions.[23, 51] Alternatively, the metal complexation reaction can be performed under milder reaction conditions by the use of Cr(CO)$_3$L$_3$ (L= CH$_3$CN, NH$_3$, pyridine) precursors, which allow the formation of (arene)Cr(CO)$_3$ complexes at much lower temperature.[23, 51] The complexation reaction of graphene and Cr(CO)$_3$(CH$_3$CN)$_3$ affords (η$^6$-graphene)Cr(CO)$_3$ complexes at temperature as low as 50 °C.[23] Room temperature complexation of arenes has been accomplished by the reaction of Cr(CO)$_3$(NH$_3$)$_3$ with an arene in the presence of BF$_3$·OEt$_2$.[51]

**4.1.2. Decomplexation of the Metal-Graphene Complexes.** The ease of metal removal in the graphitic organometallic complexes is similar to that observed previously in small molecule chemistry. The (η$^6$-arene)metal(CO)$_3$ complexes are known to undergo loss of metal in high yields at the end of a synthetic sequence.[28, 51] While the arene−metal bond can survive in a number of reaction environments, the (η$^6$-arene)Cr(CO)$_3$ complexes can be readily cleaved upon oxidation of the metal (with Ce(IV), Fe(III), I$_2$, hv/O$_2$).[51] The mildest procedure is the exposure of a solution of the complex in diethylether or acetonitrile to sunlight and air for few hours.[22] In small molecule chemistry, this method generally allows the isolation of the arene in yields that are >80%.[51]

We have recently demonstrated that for graphene this metal complexation chemistry is reversible; subjecting these metal complexes to white light in acetonitrile under an ambient atmosphere (hv/O$_2$ in CH$_3$CN) or under competitive ligand exchange conditions, (mesitylene or anisole) led to the regeneration of graphene in its almost pristine state.[22, 23]

**4.2. Atomtronics: Atomic Contacts and Interconnects via Bis-Hexahapto-Bridging of SWNTs in the Form of (η$^6$-SWNT)M(η$^6$-SWNT) Architectures.** The high electrical conductivity (10,000-30,000 S/cm) of individual single-walled carbon nanotubes (SWNTs) makes this a very promising electronic material.[52] However the high

-13-

inter-nanotube junction resistances lead to poor conductivity in thin film SWNT networks.[52, 53] Doping (charge transfer or ionic chemistry) is expected to improve the inter-CNT contacts and the conductivity of the individual CNTs, and doping with alkali metals (K) and halogens (bromine) led to an overall increase of conductivity of SWNTs by a factor of 15 to 20 at room temperature.[54] Thus particular attention has focused on the interactions of carbon materials with metals due to the ability of the metals to dope carbon materials to the point that metallic and even superconducting properties emerge.[54-56]

In a departure from previous approaches, we have recently demonstrated that covalent bis-hexahapto-bonding of transition metals at the inter-SWNT-junctions as a result of metal vapor deposition can lead to a significant improvement in the SWNT thin film conductivity and the effects are quite distinct from the non-covalent charge-transfer doping of SWNTs by alkali metals.[13]

**4.2.1. The Metal Vapor Synthesis (MVS) Technique in Constructing ($\eta^6$-SWNT)M($\eta^6$-SWNT) Oligomeric Architectures.** Timms first demonstrated the synthesis of organometallic complexes of transition metals using metal vapor synthesis (MVS) in 1969,[57] and the process has been used to synthesize a variety of compounds which incorporate metal-ligand bonds.[58] The electron beam evaporation technique was used as a source of metal vapors in 1973,[59] and the technique was shown to yield bis(arene)molybdenum complexes by condensation of molybdenum vapor with benzene, toluene, or mesitylene at 77K,[59] whereas the reaction of titanium vapor with benzene afforded extremely air-sensitive bis(benzene)titanium [$(C_6H_6)_2Ti$].[60] The MVS technique is generally employed for the synthesis of bis(arene)metal[27, 61, 62] or related (arene)−metal−(arene) oligomeric complexes, and this method allowed the synthesis of a novel triple-decker sandwich complex: ($\eta^6$-mesitylene)$_2$($\mu$-$\eta^6$:$\eta^6$-mesitylene)Cr$_2$.[63]

We have employed a derivative of this method in which bis-hexahapto-metal complexes of carbon materials are synthesized by the controlled e-beam evaporation of metal atoms onto thin films of carbon materials at room temperature with in situ measurement of properties such as the conductivity in a cryogenically pumped high vacuum chamber.[18-20]

**4.2.2. Electrical Transport Properties of the SWNT-Metal-SWNT Oligomeric Architectures.** Our recent experiments have shown that controlled e-beam evaporation of selected transition metal atoms (typical metal deposition rate = 0.2 – 0.3 Å/s) under high vacuum conditions onto SWNT thin films (MVS) leads to the formation of the bis-hexahapto-($\eta^6$)-metal bonds between SWNTs, which can function as an electrically conducting interconnect (pathway) between the side-wall benzene rings of adjacent SWNTs.[18-20] This was confirmed by in-situ transport measurements in which the conductivity of SWNT thin films (typically in the range of 2-8 nm) films was recorded as a function of the thickness of deposited metals.

The atomtronic approach, utilizing the organometallic bonding of a discrete atom, in which hexahapto-metal-bonds (atomic interconnects) are created, is distinct from the usual bulk metal contacts to graphitic materials. Atomtronics involves the insertion of an atom which is capable of hybridizing and bonding with the graphene surface without bringing about geometric rehybridization (pyramidalization) or disrupting the conjugation and which becomes a part of the electronic structure of the carbon material; note that the metal establishes a bond to the graphene surface and not a work function with a distinct Fermi level. The use of bulk metal contacts to carbon surfaces have been shown to dissipate heat, are limited by the quantum conductance at the junction, and suffers from charge transfer (doping) effects.[32, 49, 50] Separated semiconducting (SC-) SWNT thin films (thickness $t$ = 8 nm) deposited on interdigitated gold electrodes were used for our studies to make a clear differentiation between covalent hexahapto organometallic chemistry and the effects of metal to carbon charge transfer.[13, 18]

The limiting behaviors of the types of interactions between graphitic surfaces and metals can be understood by choosing metals which differ in their electronic configuration. As introduced in **Section 2** the interactions include: (a) weak physisorption (Au, Pb), (b) strong physisorption (ionic charge transfer, Li, K), (c) chemisorption with rehybridization, and (d) chemisorption, with hexahapto ($\eta^6$) complexation (Cr, Mo, W, Dy). **Figure 4** shows the results of the deposition of three distinct metal types which bring about disparate effects in the conduction behavior of the SC-SWNT films.[13, 18]

In our experiments, the deposition of gold (Au) on SWNT thin films gave a very weak response although the bulk electrical conductivity of gold is about six times that of Cr and Li. In contrast to Au, the deposition of lithium (Li) led to an increase in conductance by a factor of over 100 before reaching a maximum conductivity at about $t \approx 1.2$ nm of Li. Although the deposition of chromium (Cr) did not show such a pronounced change in the SWNT thin film conductivity, the variation in the conductivity of the SC-SWNT film showed a different dependence on the amount of metal deposited. The amount of Cr necessary to achieve the maximum effect on the conductivity of the SC-SWNT film was found to be very small, and for a Cr thickness $t < 0.05$ nm, the conductivity of the SC-SWNT film increased by more than an order of magnitude, whereas for $t = 0.05$ nm Li, there was a relatively small change in the SC-SWNT film conductivity (**Figure 4c**).[13]

These trends in the conductivity enhancements with metal deposition can be explained based on the types of interactions between the metal atoms and the graphitic surface. Deposition of Au on the SC-SWNT film leads to the parallel conductance of a weakly absorbed, continuous Au film, whereas more complex phenomena occur with the Cr and Li depositions.[13, 18] The compositions ($MC_x$), at the points where the films attained their maximum conductance, were determined to be: $t_M = 1.2$ nm for Li, giving a value of $x \approx 8$ ($LiC_8$), whereas for Cr the upper limit was reached at $t_M < 0.05$ nm, which gives $x > 180$.[13, 18]

Thin films of SWNTs consist of a network of individual SC-SWNTs and SWNT bundles, in which the transport characteristics of the films are determined by the highly resistive nature of the intercarbon nanotube junctions.[53] In earlier studies, the increase in the conductivity of SWNTs on exposure to alkali metals was attributed to charge transfer into the conduction band of the SWNTs to give a composition $KC_8$, which is reminiscent of the composition of graphite intercalation compounds.[54, 56] In contrast to such charge transfer interactions, the enhanced film conductivity due to Cr deposition is ascribed to the reduction of the internanotube resistance due to the formation of a small number of highly conductive bis-hexahapto-metal bonds that serve to electrically bridge individual SC-SWNTs, which results in the reduction in the resistance between individual SWNTs and bundles. The highly mobile chromium atoms[41] diffuse along the graphitic surfaces until they encounter a SWNT–SWNT junction or intrabundle contact with a geometry that allows formation of a bis-hexahapto-metal bond (**Figure 4d**). Interestingly, the formation of such ($\eta^6$-SWNT)Cr($\eta^6$-SWNT) interconnects at a junction is kinetically favorable because the van der Waals gap of 3.15 Å within SWNT bundles is comparable to the separation of 3.22 Å between the benzene rings in dibenzene chromium, ($\eta^6$-$C_6H_6)_2$Cr.[64] Additionally, the chromium atom possesses 6 valence shell electrons and by covalent coordination of two $\eta^6$-benzenoid ligands the resulting bis-hexahapto-bridged chromium shares a total of 18 electrons, which fills the 3d 4s 4p first row transition metal valence shell and is known to lead to a stable electronic structure.[25] In contrast, gold (Au), due to its filled outer d-orbital, is unable to participate in hexahapto complexation and cannot provide a conducting pathway at the carbon nanotube junctions.[13, 18]

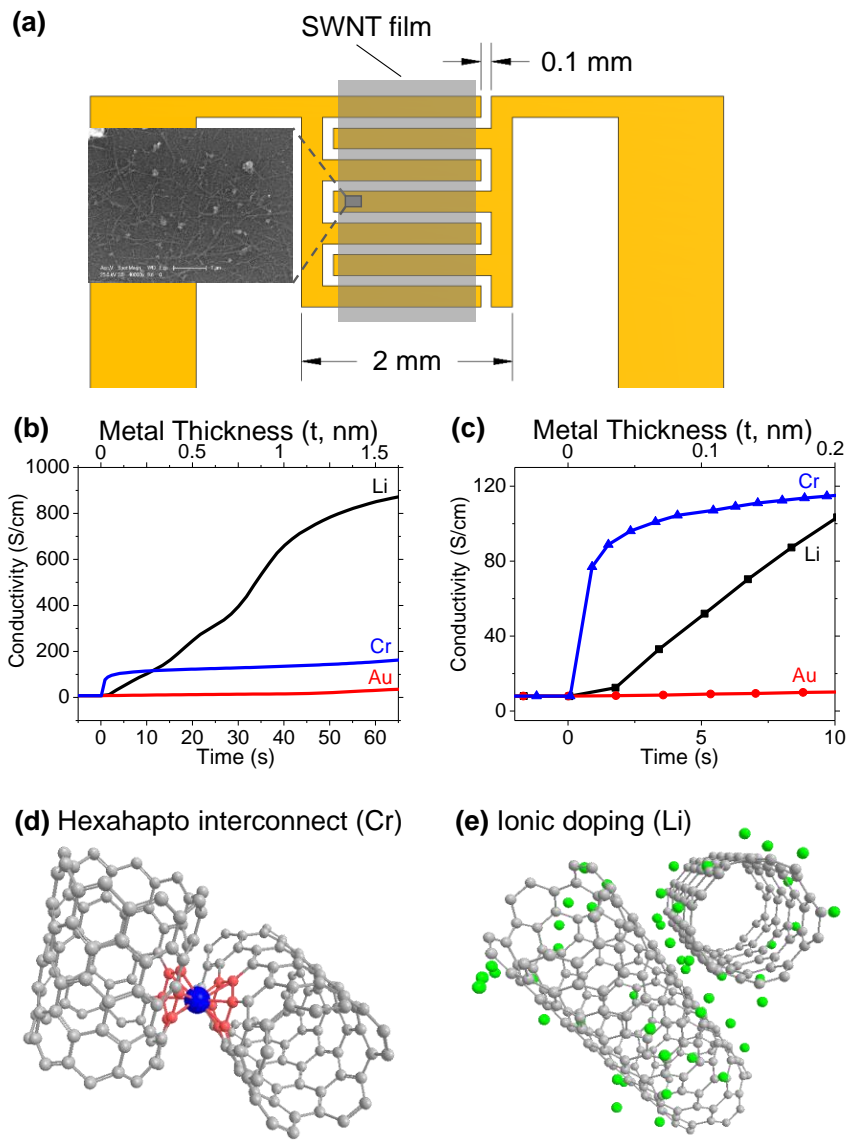

**Figure 4.** (a) Schematic diagram and scanning electron microscopy (SEM) image of a semiconducting SWNT (SC-SWNT) thin film on inter-digitated gold electrodes. (b, c) Conductivity of SC-SWNT films as a function of Li, Cr, and Au deposition. (d) Chromium atoms interconnecting adjacent nanotubes by formation of a bis-hexahapto [($\eta^6$-SWNT)Cr($\eta^6$-SWNT)] linkage, thereby reducing internanotube junction resistance. (e) Electron transfer process (doping), in which lithium atoms donate electrons to the conduction bands of the SWNTs. Adapted from ref.[18] with permission by AIP Publishing LLC. Copyright © 2012 American Institute of Physics.

The response of the conductivities of the SC-SWNT films to the deposition of first row transition metals immediately adjacent to Cr in the periodic table is also of interest due to the fact that these transition metals (M = Ti, V, Cr, Mn, and Fe) spontaneously form ($\eta^6$- $C_6H_6)_2M$ complexes by low temperature metal vapor synthesis (MVS),[25, 28] and thus are appropriate precursors for atomtronic conjugation at the SWNT junctions. It was observed in our experiments that these metals also exhibit the same sharp increase in conductivity that was found in the case of chromium; the effect of metal atom deposition on the conductivity of the SWNT thin films decreases in the order: Cr > V, Mn > Fe, Ti.[18]

The electronic structure of the chromium atoms leads to the optimum electronic structure in a bridging ($\eta^6$-SWNT)M($\eta^6$-SWNT) complex as it obeys the familiar 18-electron rule of organometallic chemistry.[25] The current studies on the hexahapto-metal complexation chemistry suggest that other transition metals, such as Ti, V, Cr, Mn, and Fe can participate in this bonding configuration (constructive rehybridization), which preserves the graphitic band structure and leads to the formation of conductive carbon nanotube interconnects via bis-hexahapto bond formation.[13, 18]

## 5. Graphene on Bulk Metal Substrates.

**5.1. Role of Metal Surfaces on CVD Growth of Graphene.** Closely-packed surfaces of (d-block) transition metals have found wide application as templates for the growth of large area, highly-ordered graphene layers of extraordinary quality.[44, 65] The reaction of metallic surfaces (such as Cu, Ni) with hydrocarbons (such as methane), in which the metallic substrate is kept at high temperature leads to the growth of graphitic sheets on the bulk metal surface.[16, 66] Careful control of temperature, pressure, quality of metal surface, growth time, and the rate of cooling can produce high quality graphene ranging from one layer to multilayers.[16, 65] The use of the high catalytic activity of low-index metallic substrates allow the growth of single graphene layer in a way that the process is self-limiting due to the inertness of the graphene surface.[44]

**5.2. Interaction between the CVD Graphene and the Underlying Metal Substrate.**
The nature of interaction (chemical bonding) between CVD grown graphene and underlying substrate is required to understand the substrate-dependent chemical reactivity, electronic structure, and the nature of the metal-graphene contact. The 2D geometry of the graphene-metal system makes it an ideal system for surface science studies of the electronic states.[44] The strength of the interaction between graphene and the underlying metal template depends on the electronic structure of the transition metal (TM) and whether the graphene lattice matches that of the metal substrate.[44] A recent study employing a combination of near-edge X-ray absorption fine structure (NEXAFS) spectroscopy, Raman spectroscopy and first-principles density functional theory (DFT) concluded that the strongest interactions occur between the commensurate (111) facets of single-crystalline metal substrates and the graphene lattice.[15]

The Graphene–Ni(111) system is lattice-matched and STM and LEED experiments show the commensurate (1 × 1) structure of carbon atoms on Ni(111).[67, 68] The relatively small distance between graphene and Ni(111) leads to significant intermixing of the valence band states of both materials and strong interaction (hybridization) between graphene $\pi$ and Ni 3d states is observed at the graphene/Ni(111) interface as seen in ARPES experiments.[44] The graphene Dirac cone is not preserved due to the strong intermixing of Ni and graphene valence band states and a large gap opens around the K point of the graphene-derived Brillouin zone. Strong covalent overlap between the metal-d and graphene-$\pi$ orbitals and hole doping of graphene upon the deposition of Ni and Co metal contacts onto graphene/$SiO_2$ was observed and the effect was significantly stronger for these metals in comparison to Cu (discussed below).[15]

Graphene on Cu(111) is also lattice-matched but it is a weakly interacting system due to the completely filled Cu d shell (unlike Fe, Co and Ni), thus the graphene-derived Dirac cone electronic structure is preserved with its position defined by the doping level. The absence of hybridization between the graphene $\pi$ and Cu 3d states around the Fermi level ($E_F$) is seen in studies of the graphene-Cu(111) interface.[15, 44] It has been suggested that the $\pi$ cloud in CVD-SLG is pinned to the Cu foil substrate after CVD growth at 1000 °C in a way that the $\pi$-electron density in graphene is no longer available for interfacial hybridization with metal d-states when metal overlayers are deposited on the CVD-graphene/Cu system. In other words, such an interaction makes the graphene top-surface electronically less accessible for chemical bonding.[15]

An interesting outcome of the recent work is the suggestion that metal deposition on graphene induces a loss of degeneracy between the A and B graphene sublattices and that spin-majority and spin-minority channels are distinctly coupled to graphene, thereby contributing to splitting of the characteristic π* resonance.[15]

**5.3. Effect of Metal Substrates on the Chemical Reactivity of Graphene.**
**5.3.1. Genesis of the Asymmetric Two Dimensional Graphene Chemistry – the Janus Graphene-Based Materials Proposition.** While it has been suggested that the growth of graphene on metal substrates pins the graphene π-electron density to the underlying metal substrate, it can lead to the possibility of anisotropic or asymmetric functionalization of graphene with the help of selective protection of individual faces of graphene in which the underlying metal substrate can act as protecting group.[15]

This concept can be extended to prepare graphene-based ultrathin two-dimensional "Janus graphene discs", in which dissimilar functionalities are covalently attached to the two faces of graphene.[69] Such anisotropic functionalization of graphene, in which the accessible face of graphene (CVD graphene on copper) is functionalized first by chemical modification techniques followed by photochemical halogenation with subsequent PMMA spin coating of the exposed surface. The PMMA/halogenated graphene layer is removed from the substrate by etching the copper with ferric sulphate, and then turned over to allow aryl radical addition using standard diazonium chemistry in aqueous solution with final removal of the PMMA with acetone leading to Janus graphene discs (**Figure 5**). This strategy takes advantage of three chemical characteristics: (i) weak pre-complexation of the bottom face of graphene with the copper substrate, (ii) the insolubility of PMMA in aqueous solution in which the diazonium chemistry is performed, and (iii) finally the solubility of the PMMA in an organic medium.

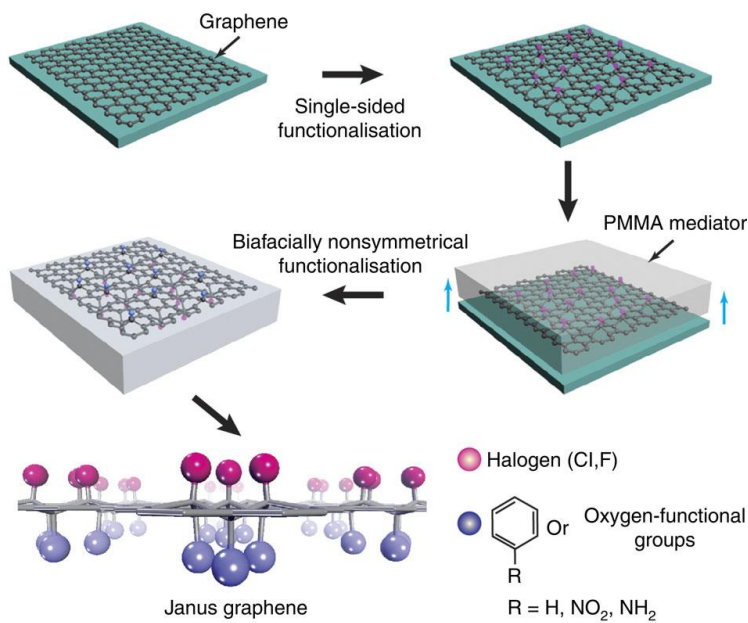

**Figure 5.** Schematic illustration of the asymmetric 2D graphene chemistry. First the accessible face of graphene (e.g. CVD graphene on copper) is chemically functionalized, then PMMA is spin coated on the single-sided functionalized graphene layer, peeling off PMMA/graphene from the copper substrate, turning over to asymmetrically modify the other side of graphene in aqueous solution with the protection of PMMA as the substrate and finally removing the PMMA to obtain thinnest Janus graphene discs. Reprinted from ref. [69] with permission by the Nature. Copyright 2013 Nature Publishing Group.

## 6. CLUSTERS

**6.1. Metal Clusters on Graphitic Surfaces: SWNT-Metal Clusters.** We recently reported the formation of chromium clusters on the outer walls of single-walled carbon nanotubes (SWNTs) by the reaction of SWNTs with chromium hexacarbonyl in dibutyl ether at 100 °C;[70] the curvature of the SWNTs and the high mobility of the chromium tricarbonyl moieties on the graphitic surfaces allows the growth of the metal clusters by decarbonylation and Cr-Cr bond forming reactions. High-resolution TEM imaging provided evidence for the presence of chromium clusters on SWNT side-walls (**Figure 6**), in contrast to traditional small molecule transition metal carbonyl cluster chemistry,[71] which suggested that carbonyl chromium complexes have little tendency to form clusters with Cr–Cr bonds and few homometallic chromium clusters have been characterized to date. The Cr–Cr bonds in these clusters may be stabilized with bridging carbonyls. Representative high-resolution TEM (HRTEM) images, given in **Figure 6**, show the

-22-

presence of small chromium clusters (2 nm – 4 nm) attached to the SWNT walls and EDX analysis confirmed the presence of chromium (**Figure 6e**).

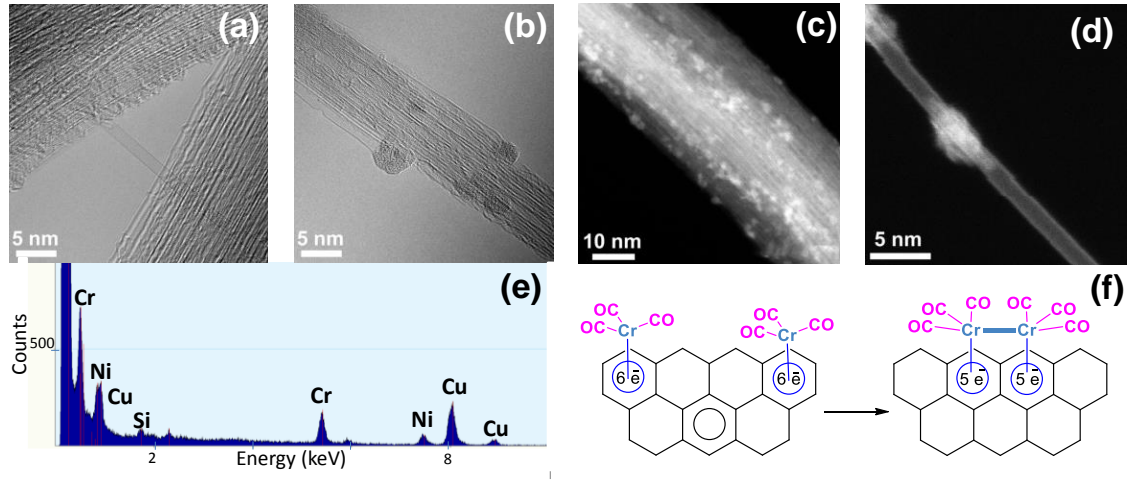

**Figure 6.** HRTEM images of (a) purified SWNTs and (b) Cr-functionalized SWNTs: ($\eta^6$-SWNT)Cr$_x$(CO)$_y$. (c) EDX measurement of ($\eta^6$-SWNT)Cr$_x$(CO)$_y$, showing the presence of Cr. (d,e) dark-field images of chromium clusters attached to SWNT bundles of ($\eta^6$-SWNT)Cr$_x$(CO)$_y$. (f) Schematic illustration of the mechanism of chromium carbonyl cluster formation on the SWNT via the graphene Clar structures. Adapted from ref. [70]. Copyright 2013 Taylor & Francis Group.

**6.2. Fermi Level Tuning of Graphene Devices with Asymmetric Metal Electrodes: Doping by Metal Contacts.** The density of states (DOS) in graphene at the Dirac point is very low and even small amounts of electron transfer at the interface to a metal significantly shifts the Fermi level. The influence of metal contacts on the work function of graphene and subsequent doping of the graphene channel has been studied by fabricating graphene devices with asymmetric metal contacts.[32] By measuring the peak position of the conductance for each pair of metal electrodes, the voltage at the Dirac point, $V_g^{Dirac}$ (V) was obtained from the gate response, and it was found that the measured shifts in $V_g^{Dirac}$ (V) were closely related to the modified work functions of the metal-graphene complexes. Within a certain bias range, the Fermi level of one of the contacts aligned with the electron band and that of the other contact aligned with the hole band. For example, the Pd-contacted device showed a positive $V_g^{Dirac}$ (V), indicating that the Fermi level in that case was aligned with the hole band of the graphene channel. In contrast, other metals (Pt, Au, and Ni) showed a negative $V_g^{Dirac}$ (V).[32] Based on these studies it was possible to classify the interactions of metals on graphene surfaces as either physisorbed (Al, Au, and Pt) or chemisorbed (Ti, Ni, Co and Pd). Further development of these ideas along the lines of previous studies of chemically functionalized graphene with Kelvin force microscopy (KFM)[32] could lead to a classification of metal interactions based on metal work functions and valence shell electronic structure.[29]

## 7. CATALYSIS.

**7.1. Reusable Hydrogenation Catalysts in Organic Synthesis.** The organometallic chemistry of carbon nanotubes, where the SWNT acts as a primary $\eta^2$–ligand or the side-wall functional groups (-COOH) acts as coordinating sites, was first explored in connection with the ability of the SWNT to coordinate Vaska's complex [Ir(CO)Cl(PPh$_3$)$_2$] and Wilkinson's catalyst complex [RhCl(PPh$_3$)$_2$].[72, 73] The practical utility and catalytic efficiency of the Rh-based SWNT–Wilkinson's complex adduct system was initially demonstrated by its usefulness in the recyclable catalytic hydrogenation of cyclohexene at room temperature.[37, 73]

On the other hand, chromium hexacarbonyl activated pristine SWNTs were used as an activated substrate in the Diels-Alder reaction.[74] Organometallic catalysis with graphene–metal complexes can provide an electronically conjugated reusable catalyst support for catalytic applications.[22]

**7.2. Proposed Catalytic Activity of Metal–Graphene Complexes in Photocatalytic Water Splitting Reaction.** Photocatalytic water splitting has attracted significant attention as a potential source for the production of hydrogen (H$_2$) - a clean and renewable energy source.[75, 76] The direct conversion of sunlight to fuel by water splitting requires at a minimum a co-catalyst working in tandem with a light absorber. Despite tremendous efforts, the development of highly active photocatalysts for splitting water at low cost still presents a challenge.[76]

In principle, when a photon with energy $E = h\nu$ matches or exceeds the band gap ($E_g$) of a semiconductor (CdS in **Figure 7c**), an electron in the valence band (VB) is excited into the conduction band (CB), leaving a positive hole ($h^+$) in the VB. The photogenerated electron-hole (e/h) pair plays a crucial role in various applications, including solar energy conversion (hydrogen production and solar photovoltaics).[75-77] However, the photogenerated electron-hole pair excited states are unstable and usually recombine to give heat, resulting in poor efficiency of the devices. Recently it was shown that the combination of graphene oxide (GO, or reduced GO) with semiconductor photocatalysts (such as $TiO_2$) can greatly enhance the photocatalyst performance, in which graphene acts as an electron acceptor to enhance the photoinduced charge transfer.[76, 77]

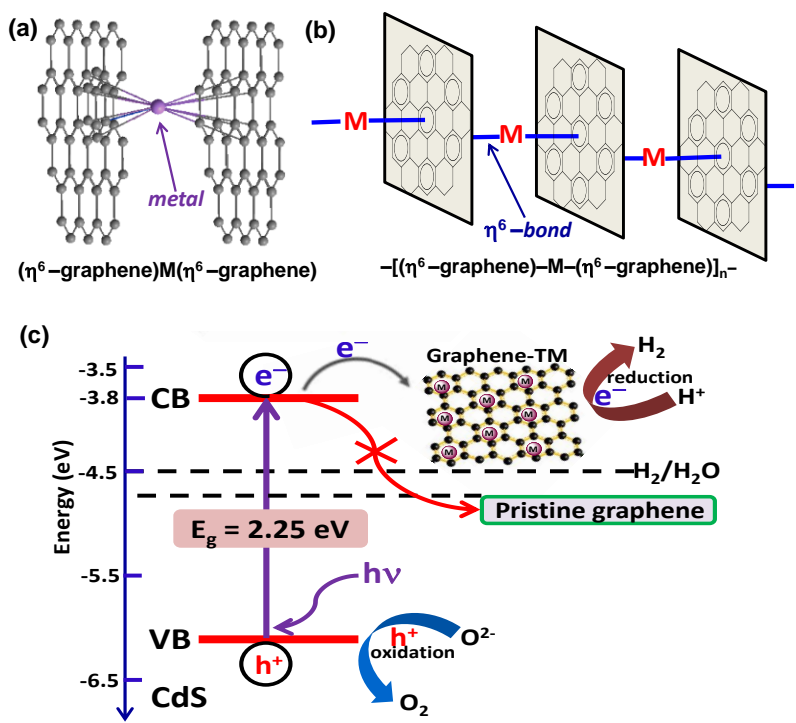

**Figure 7.** Model structures of metal−graphene catalysts in which: (a) two graphene units are connected by bis-hexahapto-bridging with a transition metal (TM), and (b) related oligomeric −[($η^6$-graphene)−metal−($η^6$-graphene)]$_n$ structures.[21, 22] (c) Proposed scheme for photocatalytic water splitting based on ($η^6$-graphene−TM)−semiconductor (CdS) hybrid (graphene−metal−semiconductor) architectures. Model distribution of metal atoms on graphene surface in a proposed graphene−metal catalyst is shown at right side, in which hexahapto-bonded metal atoms complex with Clar sextets of graphene.

We anticipate that graphene−metal−semiconductor (GMS) hybrid structures have potential as high performance novel water-splitting catalysts. While pristine graphene (work function, W ≈ 4.6 eV; $E_{VB} = E_{CB} = -4.6$ eV)[6] is not suitable for this purpose (see **Figure 7c**), transition metal deposited graphene (or related graphene−metal−graphene structures as in **Figure 7a,b**), which are reported to have a reduced value of W (due to the combined effects of charge redistribution and interfacial electrostatic dipole modifications),[78] should be an excellent choice for this purpose. This is attributed to the fact that in this case the potential of graphene/graphene⁻ is lower than the reduction potential of $H^+$ (and is closer to the CB of the semiconductor photocatalyst CdS; **Figure 7c**). This is expected to favor electron transfer from the CB of CdS ($E_{CB} = -3.8$ eV) to the vacant π* band of the graphene−metal complex, and thus to enhance the photocatalytic $H_2$-production activity (reduction of $H^+$).

Therefore, it is expected that metal deposited (or metal cross-linked) $\eta^6$-graphene−TM architectures in conjunction with appropriate semiconductors can lead to graphene−metal−semiconductor (GMS) hybrids which can function as the acceptors for photogenerated electrons for splitting of water to hydrogen under visible (or UV) light irradiation.

**7.3. Catalysts for Polymer Electrolyte Membrane Fuel Cell (PEMFC).** Identified as a clean energy source, the hydrogen operating polymer electrolyte membrane fuel cell (PEMFC) is expected to reduce dependence of fossil fuels.[79] A major issue in the widespread commercial development of the PEMFC is the high cost associated with the Pt catalyst.[79] Improving the catalyst support layers (CSL) at the anode and cathode is an important direction in this work as the CSL maintains the correct balance of reactants and products at the triple-phase interface and controls the efficiency of the catalyst.[80-82] We focused on the development of ultra-thin carbon nanotube based CSLs with ultra-low Pt loading dispersed on MWNTs, SWNTs,[81, 83, 84] and SWNT-MWNT hybrid CSLs;[80] SWNT-carbon nanofiber hybrid structures have also been reported.[82]

We have recently shown that chemically modified SWNTs with varying degrees of carboxylic acid (-COOH) functionalization are suitable for the fabrication of SWNT thin film catalyst support layers (CSLs) in PEMFCs which benchmark favorably against a number of US DOE 2017 technical targets.[36] The use of the optimum level of SWNT-COOH functionality led us to the development of a prototype SWNT-based PEMFC with total Pt loading of 0.06 $mg_{Pt}/cm^2$ - well below the value of 0.125 $mg_{Pt}/cm^2$ set as the US DOE 2017 technical target for total Pt group metals (PGM) loading. This PEMFC prototype also approaches the technical target for the total Pt content per kW of power (<0.125 $g_{PGM}/kW$) at a cell potential 0.65V: a value of 0.15 $g_{Pt}/kW$ was achieved at 80°C/22 psig testing conditions, which was further reduced to 0.12 $g_{Pt}/kW$ at 35 psig back pressure.[36] The success of this approach is attributed to the use of SWNTs of optimal bundle size together with the optimal ratio of carboxylic acid functional groups necessary to efficiently anchor and disperse the Pt catalyst nanoparticles while maintaining the essential SWNT film hydrophilicity and porosity for efficient operation of the triple phase interface.[36]

## 8. CONCLUSIONS AND PERSPECTIVES

The mobile nature of metal atoms on $\pi$-conjugated graphitic surfaces can lead to several interesting physicochemical phenomena including self-assembled metal nanoclusters with unique morphology, atomic interconnects for 3D electronics based on 1D SWNTs or 2D graphene structures (atomtronics), novel catalyst architectures, and organometallic transistor devices. This is a fertile area for new science and technology and we can expect even more interesting results in the next few years.

The detailed study of the changing morphology of metal nanostructures on graphene as a function of temperature can provide helpful information about metal thin film growth, the controlled doping of graphene together with well-behaved metal contacts. It has been suggested that the high density of the dysprosium (Dy) islands on graphene surfaces is relevant to nanometer-scale magnetism and high storage memory applications,[29] while graphene|metal|ligand systems, such as ($\eta^6$-graphene)Cr($\eta^6$-benzene),[22] may have remarkably high magnetic coupling leading to room temperature magnetic phenomena.[35]

The formation of such graphene-based organometallic systems are calculated to be accompanied by spin polarization of the graphene π-conjugated system, which can lead to "spin-valve materials and bring the realization of quantum computing one step closer".[35] A further opportunity relates to the use of these materials as electronically conjugated, reusable catalyst supports for synthetic chemistry and in materials science.[22]

Exploring the atomtronic complexes of carbon-based materials with well chosen metal atoms (M) can lead to interesting materials phenomena as exemplified by the formation of a bis-hexahapto-metal bond between the benzenoid rings of graphitic surfaces, which largely preserves the graphitic band structure (constructive rehybridization) and π-conjugation. This hexahapto ($\eta^6$-) bonding mode, unlike previous methodologies,[7, 38] leads to an enhancement in the conductivity by increasing the dimensionality of the electronic structure.[18] Such atomic, chemically formed interconnects are entirely distinct from those that depend on the physical adsorption of bulk metals, which have been labeled as a "performance killer" in the formation of metal/graphene contacts.[50]

The development of the organometallic chemistry of graphitic materials offers the potential of internally doped structures by use of transition metals, which deviate from the 18-electron rule, while still providing electrical interconnects that are applicable to the bridging of any surface containing the benzenoid ring system, including single-walled carbon nanotubes, graphenes, and other graphitic carbons.[13] The organometallic approach outlined in this Perspective may lead to new material phenomena in a number of fields, such as organometallic catalysis (for example, in fuel cells, hydrogenation and water splitting reactions),[22] memory devices,[17] high mobility organometallic transistor devices,[23] advanced energy devices, and new electronic materials of enhanced dimensionality including atomic spintronics,[35] and superconductivity.

**ACKNOWLEDGEMENTS**
This work was supported by the US National Science Foundation (NSF) under Contract DMR-1305724.